\documentclass[prl,showpacs,floatfix,twocolumn,letterpaper,footinbib]{revtex4}
\usepackage{bm,amsmath}
\usepackage{dcolumn}
\usepackage{natbib}

\newcommand{\Eref}[1]{Eq.~(\ref{#1})}
\newcommand{\tref}[1]{Table~\ref{#1}}

\begin{document}
\title{Enhanced sensitivity to variation of the fine
 structure constant and  $m_p/m_e$ in diatomic molecules}
\author{V. V. Flambaum$^{1,2}$}
\author{M. G. Kozlov$^{3,1}$}
\affiliation{$^1$School of Physics, University of New South Wales,
Sydney, 2052 Australia}
\affiliation{$^2$Institute for Advanced Study,
Massey University (Albany Campus), Private Bag 102904, North Shore MSC Auckland, New
Zealand}
\affiliation{$^3$Petersburg Nuclear Physics Institute,
Gatchina, 188300, Russia}
\date{ \today }
\pacs{06.20.Jr, 06.30.Ft}

\begin{abstract}
Sensitivity to temporal variation of the fundamental constants may
be strongly enhanced in transitions between narrow close levels of
different nature. This enhancement may be realized in a large number
of molecules due to cancelation between the ground state fine
structure $\omega_f$ and vibrational interval $\omega_v$
($\omega=\omega_f-n \omega_v\approx 0$ , $\delta \omega/\omega=K (2
\delta \alpha/\alpha +0.5 \delta \mu/\mu)$, $K \gg 1$,
$\mu=m_p/m_e$). The intervals between the levels are conveniently
located in microwave frequency range and the level widths are very
small. Required accuracy of the shift measurements is about 0.01-1
Hz. As examples, we consider molecules Cl$_2^+$, CuS, IrC, SiBr and
HfF$^+$.

\end{abstract}
\maketitle

\subsection{Introduction}\label{intro}

Theories unifying gravity with other interactions suggest temporal
and spatial variation of the fundamental "constants" in expanding
Universe. The spatial variation can explain  fine tuning of the
fundamental constants which allows humans (and any life) to appear.
We appeared in the space-time area of the Universe where the values
of the fundamental constants are consistent with our existence.
Another possible effect is dependence of the fundamental constants
on the gravitational potential which leads to the violation of local
position invariance. The strongest limits \cite{For07,FS07} are
obtained from the measurements of  dependence of atomic frequencies
on the distance from the Sun (this distance varies due to the
ellipticity of the Earth's orbit).

There are hints for the variation of the fine structure constant
$\alpha=e^2/\hbar c$, strength constant of the strong interaction,
and masses in Big Bang nucleosynthesis from quasar absorption
spectra and Oklo natural nuclear reactor data (see
e.g.\cite{MWF03,DFW04,LT04,IPV03,RBH06,MWF06,LML07}). However, a
majority of publications report only limits on possible variations
(see e.g. reviews \cite{Uza03,KFP05,Fla07}).

A very promising method to search for the variation consists in
comparison of different atomic clocks (see  recent measurements in
\cite{PTM95,Mar03,Biz05,PLS04,Biz03,Fis03,PLS05,For07}). An
enhancement of the relative variation effects happens in transitions
between very close atomic \cite{DFW99a,DFM03,ADF06,NBL04,CLN07},
molecular \cite{DeM04,DeM06,Fla06b,FK07a} and nuclear
\cite{PT03,Fla06c} energy levels.

In this paper we would like to note that very close narrow levels of
different nature exist in diatomic molecules due to cancelation
between the fine structure and vibrational intervals in the
electronic ground state. The intervals between the levels are
conveniently located in microwave frequency range and the level
widths are very small, typically $\sim 10^{-2}$~Hz. The relative
enhancement of the variation effect $K$ can exceed 5 orders of
magnitude.

\subsection{Effects of the variation and selection of molecules}
\label{selection}

The fine structure interval $\omega_f$ rapidly increases with
increase of nuclear charge Z:
\begin{align}
\label{of}
 \omega_f \sim Z^2 \alpha^2 E_H \, ,
\end{align}
where $E_H=m_e e^4/\hbar ^2$ is the atomic energy unit Hartree
 ($E_H$=2Ry=219475~cm$^{-1}$).
On the contrary, the vibration energy quantum decreases with the atomic mass:
\begin{align}
 \label{ov}
\omega_v \sim M_r^{-1/2} \mu^{-1/2} E_H
\end{align}
where  $\mu=m_p/m_e$, $m_p$ is the proton mass, $m_e$ is the
electron mass and the reduced  mass for the molecular vibration is
$M_r m_p$. Therefore, we obtain equation for the lines $Z=Z(M_r,n)$
where we have cancelation between the fine structure and vibrational
energy:
\begin{align}
 \label{o}
\omega=\omega_f - n  \omega_v \approx 0 \,, \,\, n=1,2,...
\end{align}
Using Eqs.~(\ref{of}--\ref{o}) it is easy to find dependence of the
transition frequency on the fundamental constants:
\begin{align}
 \label{do}
 \frac{\delta\omega}{\omega}=
 \frac{1}{\omega}\left(2 \omega_f \frac{\delta\alpha}{\alpha}+
 \frac{n}{2} \omega_v  \frac{\delta\mu}{\mu}\right)
 &\approx K \left(2 \frac{\delta\alpha}{\alpha}+
\frac{1}{2} \frac{\delta\mu}{\mu}\right),
\end{align}
where the enhancement factor $K= \frac{\omega_f}{\omega}$
%>mgk11/07 and 29/08>
determines the relative frequency shift for the given change of
fundamental constants. Large values of factor $K$ hint at
potentially favorable cases for making experiment, because it is
usually preferable to have larger relative shifts. However, there is
no strict rule that larger $K$ is always better. In some cases, such
as very close levels, this factor may become irrelevant. Thus, it is
also important to consider the absolute values of the shifts and
compare them to the linewidths of the corresponding transitions.
That will be done in the following sections.
%<mgk11/07 and 29/08<

Because the number of molecules is finite we can not have $\omega=0$
exactly. However, a large number of molecules have $\omega/\omega_f
\ll 1$ and $|K| \gg 1$. Moreover, an additional ``fine tuning'' may
be achieved by selection of isotopes and rotational,
$\Omega$-doublet, and hyperfine components. Therefore, we have two
large manifolds, the first one is build on the electron fine
structure excited state and the second one is build on the
vibrational excited state. If these manifolds overlap one may select
two or more transitions with different signs of $\omega$. In this
case expected sign of the $|\omega|$-variation must be different
(since the variation $\delta \omega$ has the same sign) and one can
eliminate some systematic effects\footnote{This way of control of
systematic effects was used in \cite{NBL04,CLN07} for transitions
between close levels in two dysprosium isotopes. The sign of energy
difference between two levels belonging to different electron
configurations was different in $^{163}$Dy and $^{162}$Dy.}.

Note that $\omega$ is sensitive to the variation of two most
important dimensionless parameters of the Standard Model.
The first parameter, $\alpha$, determines the strength of
the electroweak interactions. The second parameter, $\mu=m_p/m_e$,
is related to the weak (mass) scale and strong interaction scale.
Indeed, the electron mass is proportional to
the vacuum expectation value of the Higgs field (the weak scale)
which also determines masses of all fundamental particles.
The proton mass is proportional to another fundamental parameter,
the quantum chromodynamics scale $\Lambda_{QCD}$ ($m_p \approx
3\Lambda_{QCD}$). The proportionality coefficients cancel out in the
relative variation. Therefore, we are speaking about the relative
variation of a very important dimensionless fundamental parameter of
the Standard Model, the ratio of the strong to weak scale, defined
as $\delta(\Lambda_{QCD}/m_e)/(\Lambda_{QCD}/m_e)=\delta\mu/\mu$.

\begin{table}[tbh]
  \caption{Diatomic molecules with quasidegeneracy between the
  ground state vibrational and fine structures. All frequencies are
  in cm$^{-1}$. The data are taken from \cite{HH79}.}
  \label{tab1}
  \begin{tabular}{lldd}
  \hline\hline
    Molecule    &Electronic states   &\multicolumn{1}{c}{$\omega_f$}
                                                &\multicolumn{1}{c}{$\omega_v$}\\
  \hline
    Cl$_2^+$    &$^2\Pi_{3/2,1/2}   $&   645    &   645.6 \\
    CuS         &$^2\Pi             $&   433.4  &   415   \\
    IrC         &$^2\Delta_{5/2,3/2}$&  3200    &  1060   \\
    SiBr        &$^2\Pi_{1/2,3/2}   $&   423.1  &   424.3 \\
  \hline\hline
  \end{tabular}
\end{table}

In \tref{tab1} we present the list of molecules from
Ref.~\cite{HH79}, where the ground state is split in two fine
structure levels and \Eref{o} is approximately fulfilled. The
molecules Cl$_2^+$ and SiBr are particularly interesting. For both
of them the frequency $\omega$ defined by \eqref{o} is of the order
of 1~cm$^{-1}$ and comparable to the rotational constant $B$. That
means that $\omega$ can be reduced further by the proper choice of
isotopes, rotational quantum number $J$ and hyperfine components so
we can expect $K \sim 10^3-10^5$. New dedicated measurements are
needed to determined exact values of the transition frequencies and
find the best transitions. However, it is easy to  find necessary
accuracy of the frequency shift measurements. According to Eq.
(\ref{do}) the expected frequency shift is
\begin{align}
\label{do1}
 \delta\omega=2 \omega_f \left(\frac{\delta\alpha}{\alpha}+
 \frac{1}{4}\frac{\delta\mu}{\mu}\right)
\end{align}
Assuming $\delta \alpha / \alpha \sim 10^{-15}$ and $\omega_f\sim
500$~cm$^{-1}$, we obtain $\delta\omega \sim 10^{-12}$ cm$^{-1}\sim
3 \times 10^{-2}$ Hz. In order to obtain similar sensitivity
comparing hyperfine transition frequencies for Cs and  Rb one has to
measure the shift $\sim 10^{-5}$ Hz.

\subsection{Molecular ion {H\lowercase{f}F}$^+$}

The ion HfF$^+$ and other similar ions are considered by Cornell's
group in JILA for the experiment to search for the electric dipole
moment (EDM) of the electron \cite{SC04,MBD06}. In this experiment
it is supposed to trap the ions in the quadrupole RF trap to achieve
long coherence times. Similar experimental setup can be used to
study possible time-variation of fundamental constants. Recent
calculation by \citet{PMI06} suggests that the ground state of this
ion is $^1\Sigma^+$ and the first excited state $^3\Delta_1$ lies
only 1633~cm$^{-1}$ higher. Calculated vibrational frequencies for
these two states are 790 and 746~cm$^{-1}$ respectively. For these
parameters the vibrational level $n=3$ of the ground state is only
10~cm$^{-1}$ apart from the $n=1$ level of the state $^3\Delta_1$.
Thus, instead of \Eref{o} we now have:
\begin{align}
 \label{hff1}
 \omega=\omega_\mathrm{el} + \tfrac32 \omega_v^{(1)}
 - \tfrac72\omega_v^{(0)}\approx 0\,,
\end{align}
where superscripts 0 and 1 correspond to the ground and excited
electronic states. Electronic transition $\omega_\mathrm{el}$ is not
a fine structure transition and \Eref{of} is not applicable. Instead
we can write:
\begin{align}
 \label{hff2}
 \omega_\mathrm{el}=\omega_\mathrm{el,0} + q x\,,
\end{align}
where $\omega_\mathrm{el,0}$ corresponds to the value of the fine
structure constant $\alpha=\alpha_0$ and $x=\alpha^2/\alpha_0^2-1$.
The factor $q$ has been introduced in
\cite{DFW99b,DFW99a,DFM03,ADF06}
 and it appears
due to the relativistic corrections to electronic energy. In order
to calculate this factor for HfF$^+$ ion one needs to perform
relativistic molecular calculation for several values of $\alpha$,
which is far beyond the scope of this paper. However, it is possible
to make an order of magnitude estimate using atomic calculation for
Yb$^+$ ion \cite{DFM03}. According to \cite{PMI06} the
$^1\Sigma_1^+$~--~$^3\Delta_1$ transition to a first approximation
corresponds to the $6s$~--~$5d$ transition in hafnium ion. It is
well known that valence $s$- and $d$-orbitals of heavy atoms have
very different dependence on $\alpha$: while the binding energy of
$s$-electrons grows with $\alpha$, the binding energy of
$d$-electrons decreases \cite{DFW99b,DFW99a,DFM03,ADF06}.
 For the same transition
in Yb$^+$ ion Ref.~\cite{DFM03} gives $q_{sd}=10000$~cm$^{-1}$.
Using this value as an estimate, we can write by analogy with
\Eref{do}:
\begin{align}
 \label{hff3}
 \frac{\delta\omega}{\omega}
 &\approx
 \left(\frac{2q}{\omega} \frac{\delta\alpha}{\alpha}+
 \frac{\omega_\mathrm{el}}{2\omega} \frac{\delta\mu}{\mu}\right)
 \approx \left(2000 \frac{\delta\alpha}{\alpha}+
 80 \frac{\delta\mu}{\mu}\right),
\\
 \label{hff4}
 \delta\omega
 &\approx 20000~\mathrm{cm}^{-1}(\delta\alpha/\alpha+0.04 \delta\mu/\mu)\,.
\end{align}
Assuming $\delta \alpha / \alpha \sim 10^{-15}$ we obtain $\delta\omega
\sim$ 0.6 Hz.

\subsection{Estimate of the natural linewidths of the quasidegenerate
states}\label{sec_lw}

%>mgk11/07>
As we mentioned above it is important to compare frequency shifts
caused by time-variation of constants to the linewidths of
corresponding transitions.
%<mgk11/07<
First let us estimate natural linewidth $\Gamma_n$ of the
vibrational level $n$:
\begin{align}
 \label{lw1}
 \Gamma_n &= \frac{4\omega_v^3}{3\hbar c^3}|\langle
 n|\hat{D}|n-1\rangle|^2\,.
\end{align}
To estimate the dipole matrix element we can write:
\begin{align}
 \label{lw2}
 \hat{D}&=\left.\frac{\partial D(R)}{\partial
 R}\right|_{R=R_0}(R-R_0)
 \sim \frac{D_0}{R_0}(R-R_0)
 \,,
\end{align}
where $D_0$ is the dipole moment of the molecule for equilibrium
internuclear distance $R_0$. Using standard expression for the
harmonic oscillator, $\langle n|x|n-1\rangle=\left(\hbar
n/2m\omega\right)^{1/2}$, we get:
\begin{align}
 \label{lw3}
 \Gamma_n &= \frac{2\omega_v^2 D_0^2 n}{3c^3M_r m_p R_0^2}\,,
\end{align}
where $M_r m_p$ is the reduced mass of the nuclei. For the
homonuclear molecule Cl$_2^+$ $D_0=0$ and expression \eqref{lw3}
turns to zero. For SiBr molecule it gives $\Gamma_1\sim 10^{-2}$~Hz,
where we assumed $D_0^2/R_0^2\sim 0.1\,e^2$.

Now let us estimate the width $\Gamma_f$ of the upper state of the
fine structure doublet $^2\Pi_{1/2,3/2}$. By analogy with
\eqref{lw1} we can write:
\begin{align}
 \label{lw4}
 \Gamma_f &= \frac{4\omega_f^3}{3\hbar c^3}
 \left|\left\langle {}^2\Pi_{3/2}|D_1| {}^2\Pi_{1/2}\right\rangle\right|^2\,.
\end{align}
The dipole matrix element in this expression is written in the
molecular rest-frame and we have summed over final rotational
states. This matrix element corresponds to the spin-flip and turns
to zero in the non-relativistic approximation. Spin-orbit
interaction mixes ${}^2\Pi_{1/2}$ and ${}^2\Sigma_{1/2}$ states:
\begin{align}
 \label{lw5}
  \left|{}^2\Pi_{1/2}\right\rangle &\rightarrow
  \left|{}^2\Pi_{1/2}\right\rangle
  + \xi\left|{}^2\Sigma_{1/2}\right\rangle
  \,,
\end{align}
and matrix element in \eqref{lw4} becomes \cite{KFD87}:
\begin{align}
 \label{lw6}
 \left\langle {}^2\Pi_{3/2}|D_1| {}^2\Pi_{1/2}\right\rangle
 &\approx \xi \left\langle \Pi|D_1| \Sigma\right\rangle
 \sim \frac{\alpha^2 Z^2 e^3}{10(E_\Pi-E_\Sigma)}
 \,,
\end{align}
where $E_\Sigma$ is the energy of the lowest $\Sigma$-state.
Substituting \eqref{lw6} into \eqref{lw4} and using energies from
\cite{HH79} we get the following estimate for the molecules Cl$_2^+$
and SiBr:
\begin{align}
 \label{lw7}
 \Gamma_f &\sim 10^{-2}~\mathrm{Hz}\,.
 %\left\{
 %\begin{array}{ll}
 %10^{-2}~\mathrm{Hz,}&\mathrm{Cl}_2^+\,, \\
 %10^{-1}~\mathrm{Hz,}&\mathrm{SiBr}\,.
 %\end{array}\right.
\end{align}
Here we took into account that unpaired electron in SiBr molecule is
predominantly on Si (Z=14) rather then on Br (Z=35). Because of that
the fine splitting in SiBr is smaller than that of Cl$_2^+$, where
$Z=17$ (see Table~\ref{tab1}).

%>mgk11/07>
We conclude that natural linewidths of the molecular levels
considered here are of the order of $10^{-2}$~Hz. This can be
compared, for example, to the natural linewidth 12~Hz of the level
$^2D_{5/2}$ of Hg$^+$ ion, which was used in Ref.~\cite{For07}.
%<mgk11/07<

\subsection{Conclusions}

We have demonstrated that for such molecules as Cl$_2^+$ and SiBr
there are narrow levels of different nature separated by the
intervals $\lesssim 1$~cm$^{-1}$. The linewidths are on the order of
$10^{-2}~\mathrm{Hz}$. This is comparable to the accuracy, which is
necessary to reach the sensitivity $\delta \alpha / \alpha \sim
10^{-15}$ of the best modern laboratory tests. In the
high precision frequency measurements the achieved accuracy is
typically few orders of magnitude higher than the linewidth.
Therefore, molecular experiments proposed here look very promising.

Even higher sensitivity to the temporal variation of $\alpha$ can be
found in HfF$^+$ and similar molecular ions, which are considered
for the search of the electron EDM in JILA \cite{SC04,MBD06,PMI06}.
Transition amplitude between $^3\Delta_1$ and $^1\Sigma_0$ of
HfF$^+$ ion is also suppressed. Corresponding width is larger, than
for Cl$_2^+$ and SiBr because of the larger value of $Z$ and higher
frequency $\omega_f$. In Ref.~\cite{PMI06} the width of $^3\Delta_1$
state was estimated to be about 2~Hz. This width is also of the same
order of magnitude as the expected frequency shift for $\delta
\alpha / \alpha \sim 10^{-15}$. At present not much is known about
these ions. More spectroscopic and theoretical data is needed to
estimate the sensitivity to $\alpha$-variation reliably.
%>mgk11/07>
We hope that our present work will stimulate further studies in this
direction.
%<mgk11/07<
Additional advantage here is the possibility to measure electron EDM
and $\alpha$-variation using the same molecule and similar
experimental setup.

%>>>DeMille>>>
The idea of the molecular experiments proposed here is similar to
that of the Cs$_2$ experiment, which is currently under way in Yale
\cite{DeM04,DeM06,Sai05}. The main difference is that the electron
transition in Cs$_2$ goes between $^3\Sigma_u^+$ and $^1\Sigma_g^-$
and to a first approximation is independent on $\alpha$. On the
other hand the sensitivity to $\mu$ is strongly enhanced because for
Cs$_2$ the quantum number $n$ in \Eref{o} is $\sim 100$
\cite{DeM04}. Therefore, experiment on Cs$_2$ is primarily sensitive
to the time-variation of $\mu$, while this proposal allows to study
time-variation of $\alpha$. Another difference is that for the fine
structure doublets discussed here one can systematically satisfy
\Eref{o} for many molecules with properly chosen $Z$ and $M_r$. In
the HfF$^+$ ion we have ``accidental'' quasi-degeneracy of
electronic levels $^1\Sigma_0$ and $^3\Delta_1$ with very different
$q$-factors. Because of that the RF transition \eqref{hff1} is even
more sensitive to $\alpha$-variation, than transitions \eqref{o}.
Thus, the experiments discussed here are complementary to the
experiment of DeMille's group in Yale.

%%%%%%%%%%%%%%%%%%%%%%%%%%%
%%%%%%%%%%%%%%%%%%%%%%%%%%%

This work is supported by the Australian Research Council, Godfrey
fund, Russian foundation for Basic Research, grant No. 05-02-16914,
and by St.~Petersburg State Scientific Center.

%\bibliographystyle{apsrev}%
%\bibliography{../bib/julia_w,../bib/alpha,../bib/my_ref_w}

%#############
\end{document}